\documentclass[final,5p,times,twocolumn]{elsarticle}

\usepackage{amssymb}
 \usepackage{lineno}

\journal{J. of Atmospheric and Solar-Terrestrial Physics}

\begin{document}

\begin{frontmatter}



\title{Empirical analysis of the solar contribution to global mean air surface temperature change}


\author{Nicola Scafetta}

\address{Department of Physics, Duke University, Durham, NC 27708, USA.}

\begin{abstract}
The solar contribution to global mean air surface temperature change is analyzed by using  an empirical bi-scale climate model characterized by both fast and slow characteristic time responses to solar forcing: $\tau_1 =0.4 \pm 0.1$ yr, and $\tau_2= 8 \pm 2$ yr or $\tau_2=12 \pm 3$ yr. Since 1980 the solar contribution to climate change is  uncertain because of the severe uncertainty of the total solar irradiance satellite composites. The sun may have caused from a slight cooling, if PMOD TSI composite is used, to a significant warming (up to 65\% of the total observed warming) if ACRIM, or other TSI composites are used. The model is calibrated only on the empirical 11-year  solar cycle signature on the instrumental global surface temperature since 1980. The model reconstructs the major temperature patterns covering 400 years of solar induced temperature changes, as shown in recent paleoclimate global temperature records.
\end{abstract}


\end{frontmatter}

%\linenumbers

\section{Introduction}

Estimating the solar contribution to global mean air surface temperature change is fundamental for evaluating the anthropogenic contribution to climate change. This is regarded as one of the most important issues of our time. While some theoretical climate model studies [Hegerl \emph{et al.}, 2007; Hansen \emph{et al.}, 2007; IPCC, 2007] indicate that the solar variability has  little effect on climate (these studies estimate that less than 10\% of the global warming observed since 1900 is due to the Sun),  several empirical studies suggest that  large climatic variations are well synchronized with solar variations and, therefore, climate is quite sensitive to solar changes [Eddy, 1976; Hoyt  and  Schatten, 1997; White \emph{et al.}, 1997; van Loon and Labitzke, 2000;  Douglass and Clader, 2002; Kirkby, 2007; Scafetta and West, 2005, 2006, 2007, 2008; Shaviv, 2008; Eichler \emph{et al.}, 2009; Soon, 2009].

Theoretical studies rely on climate models. Two alternative  approaches are commonly used: energy balance models (EBM) [for example: Crowley, 2000; Foukal \emph{et al.}, 2004] and general circulation models (GCM) [for example, Hansen \emph{et al.}, 2007]. These models are based on the idea that climate is forced  by solar variations, volcano activity, aerosols and several greenhouse gases ($CO_2$, $CH_4$, etc). These forcings  are theoretically evaluated and used as inputs of the models. The climate sensitivities to the forcing is estimated according to the known physics. This known physics  is implemented in the models. The models contain a certain number of climate mechanisms such as water vapor feedback, cloud formation,  energy transfer, etc.  The major problem with this approach is that the physics implemented within the models may be severely incomplete.  Specifically, some key variables such as the climate sensitivity to $CO_2$ changes is severely uncertain.

For example, according the IPCC [2007] a doubling of  $CO_2$ may induce a temperature increase from 1.5 K to 4.5 K, and more. This large uncertainty is mostly due to the current poor understanding and modeling of  water vapor and cloud formation feedbacks which can have  large effects on climate  [Kirkby, 2007; Shaviv, 2008]. Indeed, significant discrepancies between climate model predictions and data are observed [Douglass  \emph{et al.}, 2007; Lean  and  Rind, 2008], and several climate mechanisms are still poorly understood, as reported by numerous scientific papers [Idso and Singer, 2009].

An alternative approach  is based on  empirical multilinear regression models. It is assumed that not all physics is known or implemented in the models.   The forcings are used as inputs of EBMs whose outputs are not the actual temperature signatures generated by the various forcings but   waveform functions that are assumed to be  proportional to such  signatures. The temperature is supposed to be a linear superposition of these rescaled output waveforms and  linear amplification coefficients are evaluated by means of a multilinear regression analysis of a given temperature record. Thus, it is assumed that:
\begin{eqnarray}\label{lokeq}
  \Delta T (t) &=& \sum_F \alpha_F S_F(t)+ N(t),
\end{eqnarray}
where: the regression coefficients, $\alpha_F$, are the linear amplification coefficient associated to a given forcing $F$; $S_F(t)$ is the output waveform generated by the chosen EBM forced with a given forcing $F(t)$; and $N(t)$ is the residual signal that is interpreted as natural climate variability.
 The above methodology has two major variants according to the particular EBM used to generate the waveforms.

 Some authors
 [North \emph{et al.}, 2004; Hegerl  \emph{et al.}, 2006, 2007] use typical EBMs. The adoption of EBMs is particularly useful if the interest focuses on local temperature records, but becomes less useful if the interest is in the global average temperature. In fact, when the EBM outputs need to be averaged on the entire globe an EBM does not perform  too much differently from a simple low pass RC-like filter  with  appropriate relaxation time responses. The relaxation time response of a thermodynamic system  is related to the heat capacity of the  system itself.
 For example, I found that   the EBM used by Crowley [2000], where the output is averaged on the entire globe, is approximately simulated with a low-pass RC-like filter with characteristic time $\tau=10$ yr, as deduced from the data published with Crowley's paper. In fact, some other authors [for example, Lockwood, 2008] use  low-pass RC-like filters with a specific characteristic time response for each forcing.

 On the contrary, other authors [Douglass and Clader, 2002; Gleisner  and Thejll, 2003; Lean  and  Rind, 2008] do not use  traditional EBMs. These authors just assume that the output waveform functions coincide with the corresponding forcing functions  with some time-lag shift. Thus, these authors use Eq. \ref{lokeq} with   $S_F(t)=F(t-\tau_F)$.

 The results of these multilinear regression model studies  are quite interesting, also because they differ significantly from each other.
 Hegerl  \emph{et al.} [2007] found a large variability of the climate sensitivity to the total solar irradiance (TSI)  changes depending on the paleoclimate temperature records that they  used. In some case these authors even found \emph{negative} values of the climate sensitivity to TSI changes which is evidently not physical because it would imply that global climate cools when TSI increases and warms when TSI decreases. Probably, the significant uncertainty present in the paleoclimate temperature reconstructions  and in the forcing functions is responsible for these ambiguous results. These results show that the multilinear regression analysis methodology is inefficient when applied to long and uncertain records.

\begin{figure}
\begin{center}
\includegraphics[angle=-90,width=22pc]{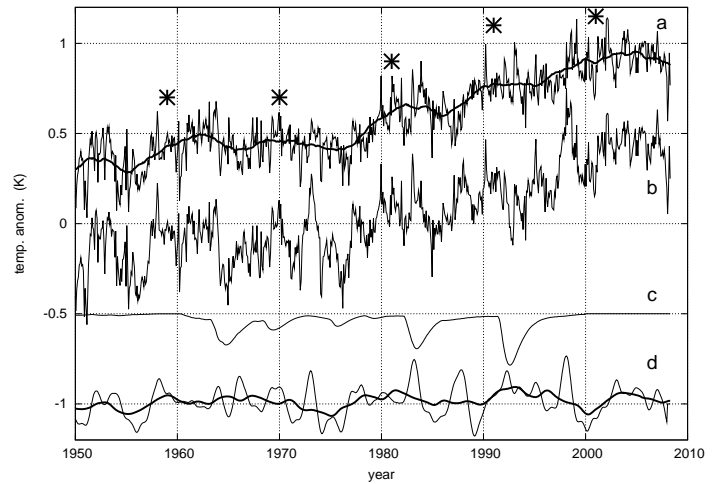}
\end{center}
 \caption{Temperature components. The curve (b) is the original global surface temperature [Brohan \emph{et al.}, 2006]. The curve (c) is the  volcano signature on the temperature  as estimated by Lockwood [2008]. The thin curve (d) is the  ENSO  signature on the temperature as estimated by Lockwood [2008]; the thick curve  is a four year moving average of the   thin curve. The thin curve (a) is the surface temperature minus the volcano and ENSO signatures plus the thick smooth curve in (d); the thick smooth curve in (a) is a four year moving average of the thin curve (a). The curves are dislocated at 0.5 K intervals  for visual convenience. The ``$\ast$'' symbols indicate the position of the TSI maxima.
 }
\end{figure}

 Lockwood [2008]  applied a nonlinear multivariate fit with several parameters on a three decades surface temperature record and   found  that the surface climate signature associated to the 11-year solar cycle has a peak-to-trough amplitude of about 0.05 K. On the contrary, Tung and Camp [2008] using similar data found a peak-to-trough solar signature amplitude of about 0.2 K.  Douglass and Clader [2002], Gleisner  and Thejll [2003],  Lean  and  Rind [2008] and several other studies [White \emph{et al.}, 1997; Scafetta and West, 2005] found that the surface climate signature associated to the 11-year solar cycle has a peak-to-trough  amplitude of about 0.1 K. Indeed, this 0.1 K solar cycle signature in the global surface temperature  appears to be the most common result among the empirical studies  [IPCC 2007, see page 674 for details], in particular since 1980. Herein, I will refer to it as the empirical estimate of  the 11-year solar cycle signature on global surface temperature since 1980.

Indeed, it is relatively easy to find this signature.  Figure 1  shows   the original global surface temperature [Brohan \emph{et al.}, 2006] (curve `b'), and the  volcano (curve `c') and the ENSO (curve `d') temperature signatures,   as recently estimated by Lockwood's  model [2008]. The curve `a' in the  figure shows  the temperature detrended of the   volcano and of the \emph{detrended} ENSO signature components. The detrended ENSO signature component is obtained by detrending the ENSO signature  of its four year moving average smooth curve, which is shown in the figure in the solid thick curve `d'. This operation does not change  the final results drastically but it is done  because the ENSO signature  may be capturing part of the  solar decadal signature on climate, so this smooth component is put back in the data before a comparison with the solar record is studied. Also Lockwook's residual signal may still contain a solar signature: therefore, it is kept in the data to avoid an inappropriate filtering.

The filtered temperature signal (curve `a' in  Figure 1) shows a clear  decadal oscillation with  a peak-to-trough  amplitude of at least 0.1 K, which is in phase with the  solar cycles. The ``$\ast$'' symbols in the figure indicate the position of the 11-year solar cycle  maxima and, on average, there is a lag-time  of about 1 year between the solar maxima and the maxima of the smooth curve `a', which fits the prediction of some EBMs [see figure 1b in North \emph{et al.}, 2004].

The peak to trough empirical amplitude regarding the 11-year solar cycle signature on global surface temperature is not reproduced by traditional GCM and EBM  estimates. North \emph{et al.} [2004] used five different EBMs and found that the climate signature associated to the 11-year solar cycle is, on average, twice  than the theoretical predictions (see their figures 1 and 4).  The climate models used by Crowley \emph{et al.} [2000], Foukal \emph{et al.} [2004] and Hansen \emph{et al.} [2007] predict an even lower solar signature on climate with a peak to trough amplitude of about 0.02-0.04 K. It is reasonable to think that  current climate models are missing important climate mechanisms that amplify the solar signature on climate, also by a large factor [Shaviv, 2008]. In fact, these models assume that the sun can alter climate only by means of direct TSI forcing while there are strong evidences that variation of direct UV radiation and cosmic rays, which affect cloud formation and change the albedo, can play a major role in climate change [Pap \emph{et al.}, 2004; Kirkby, 2007].  Thus, there are both empirical and theoretical reasons to believe that traditional climate models cannot  faithfully reconstruct the  solar signature on climate and are significantly underestimating it.

 The alternative approach that is based on multilinear regression reconstruction of climate has also some serious shortcomings. Multilinear regression analysis is very sensitive to the shape of the temperature function and to the shape of the functions used as constructors. Thus,  uncertainties  in the data and/or in the models used to construct the waveform components yield  suspicious regression coefficients, as Hegerl  \emph{et al.} [2007] found. Moreover,   multilinear regression analysis is based on the assumption of linearity and independence of the waveforms. These conditions   are not sufficiently satisfied particularly for long sequences because, for example, the solar forcing, by altering climate,  indirectly alters GHG concentrations too. Thus, GHG forcing cannot be considered independent from   solar forcing, as all studies adopting this methodology as well as most EBMs and GCMs have assumed. Consequently these models underestimate the decadal and secular solar contribution to climate change by physical construction.

 Finally, while  analyzing  shorter time series may reduce some uncertainties, when long sequences are analyzed it is  fundamental to have a physically accurate model with the correct characteristic time responses. We are left with some arbitrariness: Should we  use a model with a characteristic time response with a decadal scale,  as the EBMs used by  Crowley \emph{et al.} [2000] and North \emph{et al.} [2004] assume, or a model that uses as waveforms  the forcing functions  with short time-lag shift,  as Douglass and Clader [2002], Gleisner  and Thejll [2003] and Lean  and  Rind [2008] do?   There is a significant difference between the two approaches. In fact,  EBMs predict that the time-lag shift and the climate sensitivity to a cyclical forcing greatly increase with decreasing frequency [Wigley, 1988] and, therefore,  cannot be kept constant at all temporal scales.

 The models used by Douglass and Clader [2002], Gleisner  and Thejll [2003], Lean  and  Rind [2008]  would be severely misleading when applied to multidecadal and secular  sequences  because the amplitude of the low frequency component of the solar signature on climate would be severely underestimated. The multilinear regression approach with  fixed time-lag shifts  is more  appropriate if applied to relatively short time sequences (a few decades) and if the study is limited to evaluate the 11-year solar cycle signature on climate. In fact, if the forcing function is characterized by a unique frequency using the latter method would be approximately equivalent to using a low-pass RC-like filter model or an EBM because  a sinus-like function is transformed into a fixed time-lag shifted sinus-like function by a low-pass RC-like filter or any climate model.

 Thus, we have to conclude that both traditional climate models and multilinear regression analysis models are not completely satisfactory. The  peak to trough empirical amplitude of $0.1 ^oC$ regarding the 11-year solar cycle signature on global surface temperature seems to be sufficiently robust because it was obtained with multiple analysis methods by several authors. Moreover, while  climate responds to a given forcing with  given time responses, the climate characteristic time responses too should be  empirically measured, and not just theoretically deduced as in all above studies.

The climate characteristic time responses  were empirically measured  by Scafetta [2008a], and confirmed by Schwartz [2008], by studying the autocorrelation properties of the global surface temperature record,  assuming that the memory of its fluctuations is described by autoregressive models. The result is consistent with the  fluctuation dissipation theorem that relies on the assumption that the response of a system in thermodynamic equilibrium to a small applied force is the same as its response to a spontaneous fluctuation.

 It was found that climate is  characterized by two major characteristic time responses: one short with a time scale of a few months ($\tau_1=0.4\pm0.1$ yr); and one long with a time scale that may be as short as $\tau_2=8\pm2$ yr or, by taking into account statistical biases due to the shortness of the available  temperature record, as long as $\tau_2=12\pm3$ yr. Thus, the climate system appears to be made of two superimposed systems characterized with a fast and a slow  response to the forcings, respectively. For example, the atmosphere has a low heat capacity compared to the ocean and may be characterized by this short characteristic time response. Also cloud dynamics is relatively fast. On the contrary, the ocean has a large heat capacity that can be responsible for a decadal characteristic time response of the climate system. Moreover, other climate phenomena such as the change of albedo due to the melting of the glacials and natural forestation/desertification processes have a decadal time scale. Indeed, several studies confirm that the climate system may be characterized by both long and short response times [see references in Lockwood 2008].  The average surface temperature record should  be the result of the superposition of at least two signals generated by the two kinds of climate processes.

Herein, I propose a  model that includes the above two characteristic time responses  and I calibrate it on the   empirical estimate of  the 11-year solar cycle signature on global surface temperature since 1980, which is herein assumed to be  0.1 K.  Then, this model is  used to predict the solar signature on  climate since 1600 and this is compared with  paleoclimate temperature reconstructions.

\begin{figure}
\begin{center}
\includegraphics[angle=-90,width=22pc]{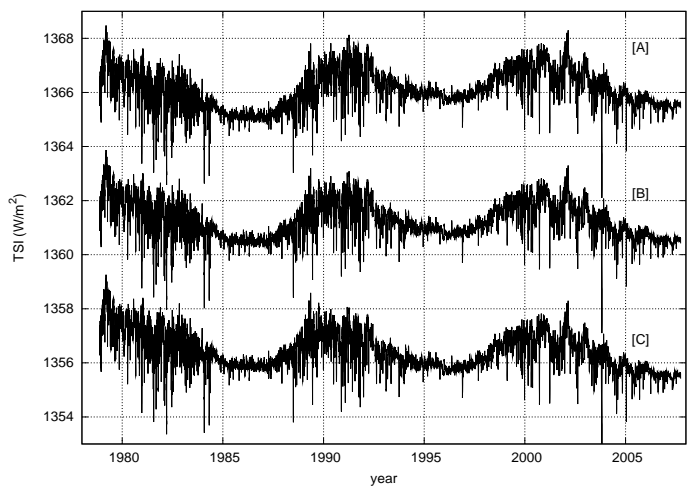}
\end{center}
 \caption{ Three alternative TSI satellite composites. The composite [B] and [C] are shifted by $-5W/m^2$ and $-10W/m^2$, respectively, from the `native scale' for visual convenience. (units of watts/meter$^2$ at 1 A.U.)
 }
\end{figure}

\section{Total solar irradiance records}

Determining how solar activity has changed on decadal and secular scales is necessary to estimate the solar contribution to climate change. Unfortunately, how  solar activity has changed in time is not known with certainty.

   Direct TSI observations started in 1978 with  satellite measurements. For the period before 1978 only TSI proxy reconstructions have been proposed (for example: Hoyt and Schatten [1997]; Lean  [2000]; Wang \emph{et al.}, [2005] and Krivova \emph{et al.} [2007]). These TSI proxy models significantly differ from each other, in particular about the amplitude of the secular trends.

Unfortunately,  TSI satellite composites since 1978 are not certain either. Two major composites have been proposed:  the PMOD TSI composite [Fr\"ohlich  and   Lean, 1998; Fr\"ohlich, 2004; Fr\"ohlich, 2006] which shows an almost  constant trend from 1980 to 2000; and  the ACRIM TSI composite [Willson and Mordvinov, 2003], which shows an increasing trend during the same period.

GCMs and EBMs adopted by the IPCC  [2007] assumed that  TSI did not change significantly since 1950 and that, consequently, the sun could not  be responsible for the significant warming observed since 1975. These estimates are based on TSI proxy models such as  those prepared by Lean  [2000] and Wang \emph{et al.} [2005]  which are apparently supported by  PMOD  [Fr\"ohlich, 2006]. However, the above TSI proxy models would be erroneous if the ACRIM TSI composite more faithfully reproduces the TSI behavior during the last decades.

The ACRIM-PMOD controversy is quite complex and, herein, a detailed discussion on this topic is not possible. A  recent work by Scafetta and Willson [2009] reopened the issue by providing  a careful analysis of the most recent TSI proxy model [Krivova \emph{et al.}, 2007] based on magnetic surface fluxes. This has been done by establishing  that a significant degradation of ERBE TSI satellite likely occurred during the ACRIM-gap (1989-1992.5), as the ACRIM team has always claimed. Moreover, Scafetta and Willson invalidated the specific corrections to Nimbus7 that the PMOD TSI composite requires and confirm the opinion of the original Nimbus7 experimental team  that no sudden increase of the Nimbus7 sensitivity occurred on September  29, 1989 [see Hoyt's statement in Scafetta and Willson, 2009]. Finally, Scafetta and Willson [2009] showed that the agreement between PMOD and the proxy reconstruction about the absence of a trend between the TSI minima in 1986 and 1996 is coincidental because a careful comparison between the proxy model and the unquestioned satellite data before and after the ACRIM-gap proves that the TSI proxy model by Krivova \emph{et al.} [2007] is missing an upward trend.

Scafetta and Willson [2009] also showed that during the ACRIM-gap Nimbus7 may have increased its sensitivity by about $0.3W/m^2$. While this error is well below that hypothesized by Fr\"ohlich ($0.86 W/m^2$) to construct PMOD TSI composite, this discrepancy is compatible with the known uncertainty of Nimbus7 whose sensors, as well as those of ERBE, were not able to make accurate self-calibrations. A direct comparison of the local trends between Nimbus7 and ACRIM1 does show  discrepancies that can be as large as $\pm 0.3 W/m^2$ [Scafetta, 2009].

Consequently,  herein I assume three TSI composites  [Scafetta, 2009] that  approximately  cover the entire range of possible TSI satellite composites. The composites are constructed  by taking into account overlapping regions and continuity at the merging dates. The composite [A] assumes that during the ACRIM-gap  the Nimbus7 record is accurate; the composite [C] assumes that during the ACRIM-gap the  ERBE record is  accurate;  the composite [B] is just the arithmetic average between [A] and [C]. Note that the  ACRIM composite is approximately  between the composite [A] and [B], while the PMOD composite is approximately reproduced by the composite [C]. All configurations between the composites [A] and [C] are ideally possible according to the published TSI  satellite records.  As Figure 2 shows, the possible range of TSI difference between the two solar minima in 1986 and 1996 is $\Delta I \approx 0.3\pm0.4 W/m^2$.

   A secular TSI record  can be  obtained, even if imperfectly,  by merging a TSI secular proxy reconstruction (here I use  the most recent TSI proxy reconstruction by  Solanki's team [Krivova \emph{et al.}, 2007]) with the three  TSI satellite composites shown in Figure 2. The 1980-1990 TSI mean values are used for this merging. The three composed TSI records are shown in Figure 3 and indicated with [A], [B] and [C].

\begin{figure}
\begin{center}
\includegraphics[angle=-90,width=22pc]{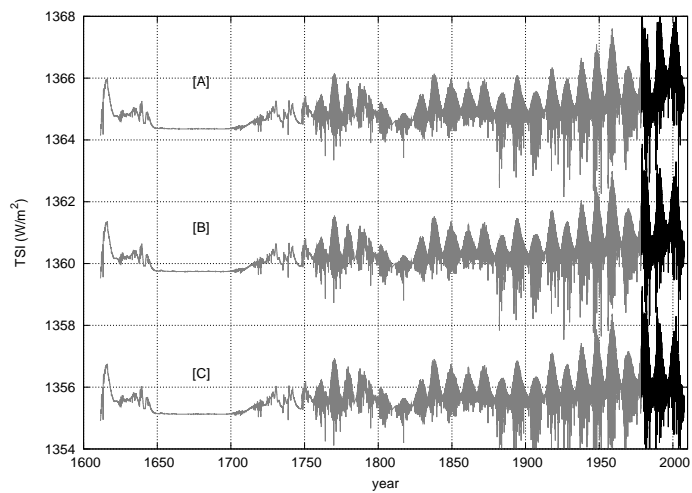}
\end{center}
 \caption{
  Three alternative secular TSI  records constructed by merging the TSI proxy reconstruction by Solanki's team [Krivova \emph{et al.}, 2007]  with the three alternative TSI satellite composites shown in Figure 1. The composite [B] and [C] are shifted by $-5W/m^2$ and $-10W/m^2$, respectively, from the `native scale' for visual convenience. (units of watts/meter$^2$ at 1 A.U.)
 }
\end{figure}

\section{An empirical climate model with short and long  characteristic time responses}

 As explained in the Introduction, climate appears to be  characterized by at least  two  major characteristic time constants by both theoretical and empirical findings [Scafetta, 2008a; Schwartz, 2008]:  a short characteristic time constant,  $\tau_1=0. 4 \pm 0.1$ yr, and a decadal one from $\tau_2=8\pm2$ yr to $\tau_2=12\pm3$ yr. Consequently, the  signature of a given forcing $F(t)$ on the global surface temperature, $\Delta T_F(t)$, is  the superposition of at least two major signals: one generated by the processes with a short time response  and the other  generated by those processes with a long   time response to a forcing. In the case of the solar forcing we should write:
\begin{eqnarray}\label{newm}
\Delta T_S(t)&=& \Delta T_{1S}(t)+\Delta T_{2S}(t),
\end{eqnarray}
where
\begin{eqnarray}\label{newii1}
   \frac{d\Delta T_{1S}(t)}{dt} &=& \frac{k_{1S} \Delta I(t)-\Delta  T_{1S}(t)}{\tau_{1S}}, \\\label{newii2}
   \frac{d\Delta T_{2S}(t)}{dt} &=& \frac{k_{2S} \Delta I(t)-\Delta  T_{2S}(t)}{\tau_{2S}}.
\end{eqnarray}
The parameters $\tau_{S1}$ and $\tau_{S2}$ are the short and long climate  characteristic time responses, and $k_{1S}$ and $k_{2S}$ are the average equilibrium climate sensitivities   of the two kind of  processes. With appropriate parameters, the above model  simulates the performance of any energy balance models, as those used by North \emph{et al.} [1981] and Crowley \emph{et al.} [2000].

The above equations assume that the TSI record is used  as a \emph{proxy} for the overall climate sensitivity to solar changes. Thus, the parameters $k_{1S}$ and $k_{2S}$ do not have  the  meaning of climate sensitivity to TSI variation as assumed in the EBMs and GCMs, but they have a meaning of climate sensitivity to ``total solar activity" variation in TSI  units. The difference is fundamental because climate is also altered  by solar changes alternative to TSI changes. The empirical methodology I propose to evaluate $k_{1S}$ and $k_{2S}$  is summarized in four steps:

1) We are interested in evaluating the solar signal on climate.   This is determined by Eq. \ref{newm} which depends on  four parameters, $k_{1S}$, $\tau_{1S}$,  $k_{2S}$ and $\tau_{2S}$.

2) For the fast process I set  $\tau_{1S}=0.4$ yr as empirically measured [Scafetta, 2008a], and $k_{1S}=0.053 ~K/Wm^{-2}$. The justification for the latter value is that the climate system has no sufficient time to physically, chemically and biologically evolve, that is, I assume that on this short time scale the chemical composition of the air, the ice cover, the forest cover and all other major physical climate variables do not change much. In this situation the climate sensitivity can be approximately calculated as
\begin{equation}\label{}
    k_{1S}=\frac{dT}{dI}=\frac{T}{4I}=0.053 ~K/Wm^{-2},
\end{equation}
 where the average Earth surface temperature is $T=289 ~K$ and the average TSI is $I=1365 ~W/m^2$. The above value is easily calculated by differentiating the energy balance law,
 \begin{equation}\label{sbl}
   T^4=\frac{g_T(1-a_T)I}{4\sigma }=h(T)I,
\end{equation}
 where $\sigma$ is the Stefan-Boltzmann constant. The albedo ``$a_T$'' as well as  any additional climate function ``$g_T$'', which is needed to correct the black-body Stefan-Boltzmann law to make it compatible with the surface Earth's climate system,  are assumed to be \emph{constant} on very short time scales.  The function ``$g_T$'' would include all feedback mechanisms, emissivity, dispersion of energy in the ocean, etc.  Thus, for short time scales it is assumed that the function $h(T)=g_T(1-a_T)=const$.   Note that also Lockwood [2008, table 1] found that a climate sensitivity of $k=0.052 ~K/Wm^{-2}$ to solar changes is associated to  a   monthly characteristic time response.

3) The long characteristic time constant of climate is set to be as short as $\tau_{S2}=8$ yr or, alternatively, as long as $\tau_{S2}=12$ yr, as Scafetta [2008a] and Schwartz [2008] have found.

4) The last free parameter  $k_{2S}$ is determined by imposing that  the peak-to-trough amplitude of the global climate response to the 11-year solar cycle is about 0.1 K near the surface after 1980, as reported by the  IPCC [2007, page 674 for details], as found by several empirical studies discussed in the Introduction and as shown in  Figure 1.

A reader should be careful here because s/he might mistakenly believe that I am not taking into account the other climate forcings such as volcano and GHG  forcings.  The four  parameters  ($k_{1S}$, $\tau_{1S}$,  $k_{2S}$ and $\tau_{S2}$) of the model have been set by using findings that do take into account the other forcings. For example, $k_{1S}$ is directly calculated from the energy balance law; $\tau_{S1}$ and $\tau_{S2}$ are measured from the autocorrelation properties of temperature residual  after detrending a theoretical effect of all forcings  [Scafetta, 2008]; the peak-to-trough amplitude of the global climate signature to the 11-year solar cycle is found to be about 0.1 K near the surface after removing the volcano, GHG+aerosol  and ENSO signal [Douglass and Clader, 2002; Lean  and  Rind, 2008]. These values have been estimated with the best data available since 1980. Thus, by using such values as  constraints of the model I am taking into account the other forcings too, but implicitly, through the constraints.

\begin{figure}
\begin{center}
\includegraphics[angle=-90,width=22pc]{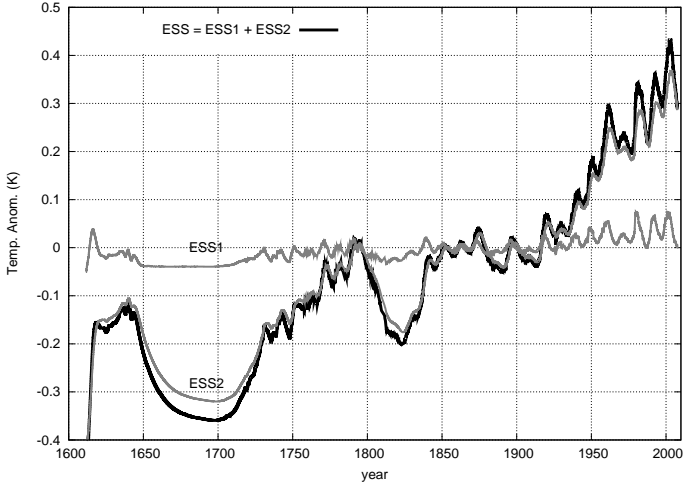}
\end{center}
 \caption{An example of ESS curves. ESS1 curve  shows that the empirical solar signature on climate obtained with Eq. (\ref{newm3}) with $\tau_1=0.4$ yr and $k_{1S}=0.053 K/Wm^{-2}$. ESS2 curve   shows that the empirical solar signature on climate obtained with Eq. (\ref{newm4}) with $\tau_2=12$ yr and $k_{2S}=0.41 K/Wm^{-2}$. ESS (black line) shows the superposition of ESS1 and ESS2. The model is forces with the TSI reconstruction [B] shown in Figure 3. The figure shows temperature anomalies relative to the 1895-1905 average.
 }
\end{figure}

Eq. \ref{sbl} cannot be used directly because it is not  known how the climate function $h(T)$ changes as function of the temperature. Thus, I use the 0.1 K solar cycle signature near the surface as a natural oscillatory output that enables us to determine the value of the climate modulation transfer function to the 11-year solar variation, and from this I can determine $k_{2S}$, as explained below.

The  climate sensitivity to the 11-year solar cycle is about $Z(11)=0.11 \pm 0.02 ~K/Wm^{-2}$
[Douglass and Clader, 2002; Scafetta and West, 2005], that is,
an 11-year solar cycle of about $1 ~W/m^2$ causes a peak to trough cycle of about  $0.1 ~^oC$ on the global surface. Thus, Eq. (\ref{newm}) is solved by the  system:
\begin{eqnarray}
\label{newm1}
  Z(11) & \approx & Z_1(11) + Z_2(11)\\
\label{newm3}
  k_{1S} &=& Z_1(11) \sqrt{1+\left(\frac{2\pi \tau_1}{11}\right)^2} \\\label{newm4}
  k_{2S} &=& Z_2(11) \sqrt{1+\left(\frac{2\pi \tau_2}{11}\right)^2},
\end{eqnarray}
where `11' refers to the  period of the 11-year solar cycle, and $Z_1(11)$ and  $Z_2(11)$ are the climate sensitivities to the 11-year solar cycle relative to climate processes with fast and slow characteristic time responses, respectively.

The above system can be easily solved by evaluating first $Z_1(11)$, then $Z_2(11)$ and, finally, $k_{2S}$. By using  Eq. (\ref{newm3}) with  $\tau_1=0.4$ yr and $k_{1S}=0.053 ~K/Wm^{-2}$, it is found:  $Z_1(11) \approx 0.051 ~K/Wm^{-2}$ and $Z_2(11) \approx 0.059 ~K/Wm^{-2}$. The former value is about 46\% of the empirical climate sensitivity to the 11-year solar cycle $Z(11)=0.11 \pm 0.02 ~K/Wm^{-2}$. Thus, the peak-to-trough amplitude of the global climate response to the 11-year solar cycle associated to the fast processes is about 0.046 K near the surface against the empirical value of  0.1 K. If  $\tau_2 \approx 8$ yr, then $k_{2S}\approx0.28 ~K/Wm^{-2}$; if  $\tau_2 \approx 12$ yr, then $k_{2S}\approx0.41 ~K/Wm^{-2}$. With the above values the overall  climate sensitivity to solar changes at equilibrium is $k_{S}\approx k_{1S} + k_{2S}\approx0.33 ~K/Wm^{-2}$ and $k_{S} \approx 0.46 ~K/Wm^{-2}$, respectively. For convenience, the parameters used in the model  are summarized in Table 1.

\section{The empirical solar signature on climate}

Figure 4 shows the empirical solar signature (ESS) on climate under the assumptions and  scenarios  mentioned above.  ESS1 and ESS2 are the empirical solar signatures induced by the climate processes with fast, $\tau_{S1}$, and slow, $\tau_{S2}$, characteristic time responses, respectively. The curves  ESS1, ESS2 and their superposition are shown in the particular case in which $\tau_2=12$ yr and the TSI record [B] is adopted.

The curve ESS1 shows a very small multidecadal and secular variability.
According to  ESS1,  the sun has induced about $+0.05K$ of warming from 1900 to 2000. Because the global warming observed since 1900 has been about $0.8K$, the sun would have caused about 6\% of the observed warming through those mechanisms that respond quickly to changes in TSI. This result does not differ significantly from the IPCC [2007] average estimates obtained with the current climate models.

However,  the small secular variability shown in ESS1 is insufficient to explain the preindustrial secular variability of the temperature, as estimated by the paleoclimate proxy temperature reconstructions  [North  \emph{et al.}, 2006]. These paleoclimate temperature reconstructions show that the preindustrial temperature variability (from 1600 to 1900) ranges from $0.05K$ to $0.5K$, while  ESS1 in Figure 4 shows a preindustrial temperature variability of about $0.04K$. The small GHG variations observed before 1900 and volcano activity could fill the difference only in the extreme case that the preindustrial climate is almost stable, as in the  \emph{hockey stick} temperature graph by Mann \emph{et al.} [1999].

However, the hockey stick temperature graph   is unlikely because the Medieval Warm Period and the Little Ice Age, which occurred during a maximum and a minimum of solar activity, respectively, are supported by numerous historical facts  [Hoyt  and Schatten, 1997] and  data from several regions of the Earth [Loehle and McCulloch, 2008]. A global warming between the periods 1650-1700 and 1850-1900, at least equal to the average among the paleoclimate temperature reconstructions and more, ($0.3K<\Delta T<0.5K$), can be considered  more realistic. Thus, unless most of the paleoclimate proxy temperature reconstructions  are severely erroneous and the historical evidences are severely misleading, Figure 4 further stresses that the processes generating ESS1 solar signature on climate alone cannot explain climate change on a secular scale.

\begin{table}
  \centering
\begin{tabular}{|c|c|c|}
  \hline
   & Case 1 &  Case 2 \\\hline
  $\tau_1$ & $0.4~yr$ & $0.4~yr$ \\\hline
  $\tau_2$ & $8~yr$ & $12~yr$ \\\hline
  $k_{1S}$ & $0.053 ~K/Wm^{-2}$ & $0.053 ~K/Wm^{-2}$ \\\hline
  $k_{2S}$ & $0.28 ~K/Wm^{-2}$ & $0.41 ~K/Wm^{-2}$ \\\hline
  $k_{S}$ & $0.33 ~K/Wm^{-2}$ & $0.46 ~K/Wm^{-2}$ \\\hline
  $Z_1(11)$ & $0.051 ~K/Wm^{-2}$ & $0.051 ~K/Wm^{-2}$ \\\hline
  $Z_2(11)$ & $0.059 ~K/Wm^{-2}$ & $0.059 ~K/Wm^{-2}$ \\\hline
  $Z(11)$ & $0.11~ K/Wm^{-2}$ & $0.11~ K/Wm^{-2}$ \\
  \hline
\end{tabular}
  \caption{The parameters used in the  empirical bi-scale climate model Eqs. \ref{newm} for  two alternative values of $\tau_2$. }\label{}
\end{table}

Figure 4 shows that most of the multidecadal and secular variability is captured by ESS2 curve. The climate processes with a slow time response to TSI changes are necessary to correctly interpret the multidecadal and secular climate changes, also since 1980. This large sensitivity is mostly due to the fact that the climate function $h(T)$, that appears in Eq. \ref{sbl}, is temperature dependent. On large time scales its temperature dependency  cannot be neglected because it has a large impact on climate.

 For example, if a $1 ~W/m^2$ TSI increase  causes at equilibrium an increase of the function $h(T)$ by just 0.5\% it is easy to calculate that the climate sensitivity to TSI change would be $k_S =\delta T/\delta I =0.41 K/Wm^{-2}$. This value is  eight times larger than  $k_{1S}=0.053 ~K/Wm^{-2}$ as it was calculated by assuming that the  function $h(T)$ is constant. This sensitivity is easily calculated with the following equation:
 \begin{equation}\label{sensi}
   \left[\frac{T_0+\delta T}{T_0}\right]^4 =\frac{h(T_0+\delta T)}{h(T_0)}~\frac{I_0+\delta I}{I_0},
  \end{equation}
  where $T_0=289~K$ and $I_0=1365~W/m^2$.
 So, even small temperature dependency of the energy-balance equation parameters, such as the albedo and the emissivity, can  amplify the climate sensitivity to solar changes by a large factor, as also Shaviv [2008] noticed.

Figure 4 also shows that the decadal variability associated to the 11-year solar cycle  derives almost evenly from both kind of climate processes with fast and slow characteristic time responses, as reproduced in the ESS1 and ESS2 curves.

\begin{figure}
\begin{center}
\includegraphics[angle=-90,width=22pc]{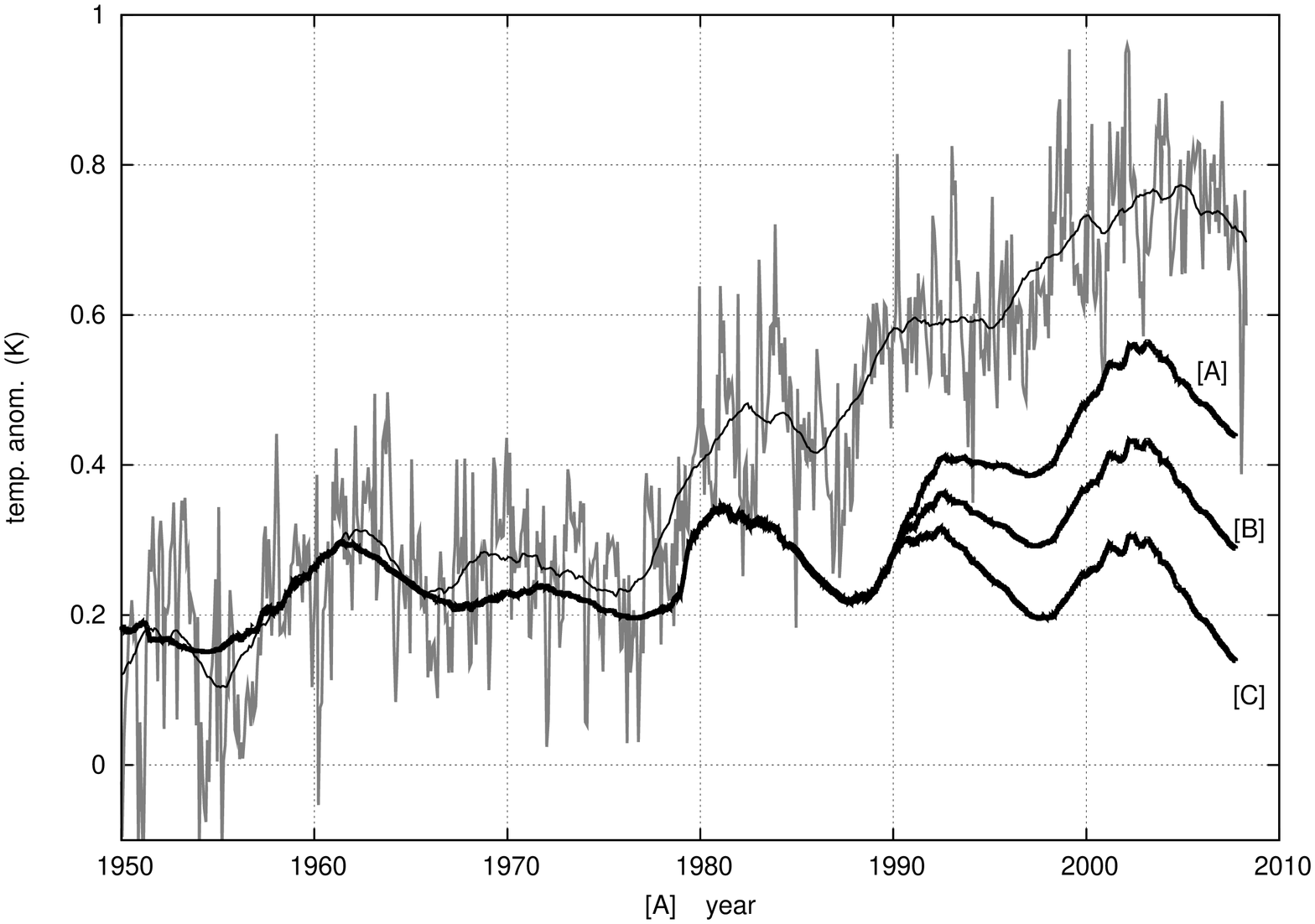}
\includegraphics[angle=-90,width=22pc]{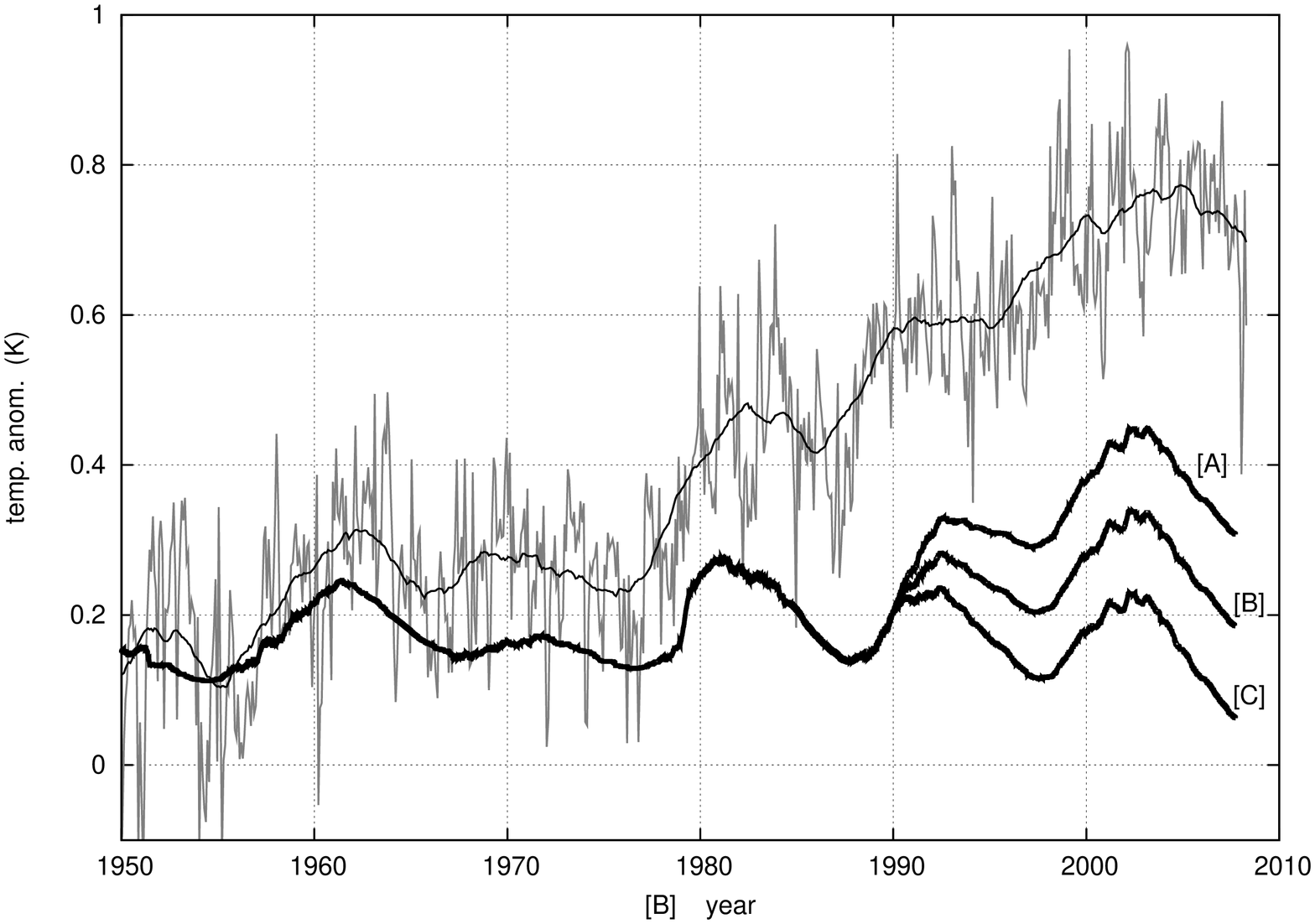}
\end{center}
 \caption{
ESS curves against the filtered global surface temperature record   shown in the curve (a) of Figure 2. [A] The ESS curves  are obtained with $\tau_1=0.4$ yr, $\tau_2 = 12$ yr and $k_{1S}=0.053 K/Wm^{-2}$ $k_{2S}=0.41 K/Wm^{-2}$. [B] The ESS curves  (blak) are obtained with $\tau_1=0.4$ yr, $\tau_2 = 8$ yr and $k_{1S}=0.052 K/Wm^{-2}$ $k_{2S}=0.28 K/Wm^{-2}$. The model is forces with the TSI reconstruction [A], [B] and [C] shown in Figure 3. The figure shows temperature anomalies relative to the 1895-1905 average.
 }
\end{figure}

Figures 5A and 5B show the ESS curves obtained by superimposing  ESS1 and ESS2 curves under different  assumptions and scenarios against the  global surface temperature record since 1950 detrended of the volcano and ENSO signatures (see curve `a' in Figure 1).  Figures  5A and 5B use  $\tau_2 = 12$ yr and $\tau_2 = 8$ yr, respectively. According to the figures there would be no significant solar induced warming since 1950 if the TSI record [C] is used; in this case the trend would be slightly negative since 1980, as also Lockwood [2008] found using  PMOD. Instead, the sun can induce as much as 65\% of the observed warming  if the TSI record [A] is used. On average the sun may have induced a significant warming  since 1950 and since 1980, respectively.  Figures 5A and 5B show also the good correspondence between the 11-year solar cycle signatures and the cycles observed in the four year moving average smooth curve of the filtered global surface temperature record.

Finally, Figure 6 shows a four century comparison between two ESS curves and a paleoclimate temperature reconstruction [Moberg \emph{et al.}, 2005] from 1600 to 1850 and the global surface temperature record since 1850. The two ESS curves shown in the figure are those with the highest secular variability (curve \#1: $\tau_2 = 12$ yr and with TSI [A]), and with the lowest secular variability (curve \#2: $\tau_2 = 8$ yr and with TSI [C]), respectively. The figure shows that the ESS signals reproduce quite well the cooling and warming patterns observed in the temperature record for four centuries, in particular with the ESS curve \#1. Since 1980 there is a wide fork whose extremes  depend on which TSI satellite composite is used. If the TSI composite [C] is used, the sun would have caused just a slight cooling. If the TSI composite [A] is used, the sun would have caused a significant warming. Thus, on average it is not unlikely that the sun  has induced a significant warming since 1980 as it was inferred by Scafetta and Willson [2009].

\begin{figure}
\begin{center}
\includegraphics[angle=-90,width=22pc]{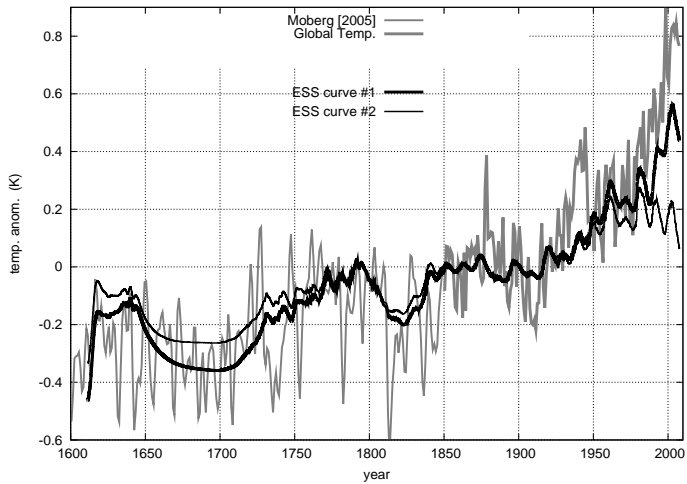}
\end{center}
 \caption{
ESS curves against a paleoclimate temperature reconstruction [Moberg \emph{et al.}, 2005] from 1600 to 1850 (thin grey line) and global surface temperature record since 1950 [Brohan \emph{et al.}, 2006] (thick black line). The ESS (thick grey) curve  is obtained with $\tau_1=0.4$ yr, $\tau_2 = 12$ yr and $k_{1S}=0.053 K/Wm^{-2}$ $k_{2S}=0.41 K/Wm^{-2}$ with the model forced with  the TSI reconstruction [A]. The ESS (thin black) curve  is obtained with $\tau_1=0.4$ yr, $\tau_2 = 8$ yr and $k_{1S}=0.053 K/Wm^{-2}$ $k_{2S}=0.28 K/Wm^{-2}$ with the model forced with  the TSI reconstruction [C].
 }
\end{figure}

\section{Conclusion}

Herein I have analyzed the solar contribution to global mean air surface temperature change. A comprehensive interpretation of multiple scientific findings indicates that the contribution of solar variability to climate change is significant and that the temperature trend since 1980 can be large and  upward. However, to correctly quantify the solar contribution to the recent global warming it is necessary to determine the correct TSI behavior since 1980. Unfortunately, this cannot be done with certainty yet. The PMOD TSI composite, which has been used by the IPCC and most climate modelers, has been found to be based on arbitrary and questionable assumptions [Scafetta and Willson, 2009]. Thus, it cannot be excluded that TSI increased from 1980 to 2000 as claimed by the ACRIM scientific team.

The  IPCC [2007] claim that the solar contribution to climate change since 1950 is negligible may be based on wrong solar data in addition to the fact that the EBMs and GCMs there used  are missing or poorly modeling several climate mechanisms that would significantly amplify the solar effect on climate. When  taken  into account the entire range of possible TSI satellite composite since 1980,  the solar contribution to climate change ranges from  a slight cooling to a significant warming, which  can be as large as  65\% of the total observed global warming.

The above wide range strongly contrasts with some recent estimates such as those found by Lockwood [2008], who  calculated that the solar contribution to global warming is negligible since 1980: the sun could have caused  from a -3.6\% using PMOD to a +3.1\% using ACRIM. In fact,  Lockwood's model is approximately reproduced by the ESS1 curve that refers to the solar signature on climate as produced only by those  processes characterized with a short time response to a forcing. Indeed, the characteristic time constants that Lockwood found with his complicated nonlinear multiregression analysis are all smaller than one year (see his table 1) and the climate sensitivity to TSI that he found is essentially equal to my $k_{1S}$! Likely, Lockwood's model was unable to detect the climate sensitivity to solar changes induced  by those climate mechanisms that have  a decadal characteristic time response to solar forcing: mechanisms that must be present in nature for physical reasons. As proven above, these mechanisms are fundamental to properly model the decadal and secular trends of the temperature because they yield  high climate sensitivities to solar changes.

Analogously, my findings  contrast with Lean and Rind [2008], who estimated that the sun has caused less than 10\% of the observed warming since 1900. The model used by Lean and Rind, like Lockwood's model, is not appropriate to evaluate the multidecadal solar effect on climate. In fact, Lean and Rind do  not use any  EBM to generate the waveforms they use in their regression analysis. These authors   assume that the temperature is just the linear superposition of the forcing functions with some fixed time-lags. They also ignore ACRIM TSI satellite composite. While Lean and Rind's method may be sufficiently appropriate for determining the 11-year solar cycle signature on the temperature records there used, the same method  is not appropriate on multidecadal scales because climate science predicts that time-lag and the climate sensitivity to a forcing is frequency dependent.  Consequently, as Lockwood's model, Lean and Rind's model too misses the larger sensitivity  that the climate system is expected to present to solar changes at the decadal and secular scales.

I have shown that the processes with a long  time response to climate forcing  are fundamental to correctly understanding the decadal and secular solar effect on climate (see ESS2 curve). With simple calculations it is possible to determine that if the climate parameters  (such as the albedo and the emissivity, etc.) change slowly with the temperature, the climate sensitivity to solar changes is largely amplified as shown in Eq. \ref{sensi}.

This finding  suggests that the climate system is hypersensitive to the climate function $h(T)$ and even small errors in modeling $h(T)$ (for example, in modeling how the albedo, the cloud cover, water vapor feedback, the emissivity, etc. respond to changes of the temperature on a decadal scale) would yield the climate models to fail, even by a large factor, to appropriately determine the solar effect on climate on decadal and secular scale. For similar reasons, the models also present a very large uncertainty in evaluating the climate sensitivity to changes in $CO_2$ atmospheric concentration [Knutti and Hegerl, 2008]. This large sensitivity of the climate equations to physical uncertainty makes the adoption of traditional EBMs and GCMs quite problematic.

 About the result depicted in Figure 6, the ESS curve has been evaluated by calibrating the proposed empirical bi-scale model  only by using the information deduced: 1) by the instrumental temperature and the solar records since 1980   about the 11-year solar signature on climate; 2) by  the findings by Scafetta [2008a] and Schwartz [2008] about the long and short characteristic time responses of the climate as deduced with autoregressive models. The paleoclimate temperature reconstructions were not used to calibrate the model, as done in Scafetta and West [2007]. Thus, the finding shown in Figure 6 referring to the preindustrial era  has also a \emph{predictive} meaning, and implies that climate had  a significant preindustrial variability which is incompatible with a \emph{hockey stick} temperature graph.

\textbf{Acknowledgment:} NS thanks the Army Research Office for research
support (grant W911NF-06-1-0323).

\newpage

\end{document}